# Dynamic motion trajectory control with nanoradian accuracy for multi-element X-ray optical systems via laser interferometry

Authors: Sina M Koehlenbeck[1*], Lance Lee[2], Mario D Balcazar[2], Ying Chen[2], Vincent Esposito[2], Jerry Hastings[2], Matthias C Hoffmann[2], Zhirong Huang[2,3], May-Ling Ng[2], Saxon Price[2], Takahiro Sato[2], Matthew Seaberg[2], Yanwen Sun[2], Adam White[2], Lin Zhang[2], Brian Lantz[1], Diling Zhu[2*]

[1] Edward L. Ginzton Laboratory, Stanford University, Stanford, California 94305, USA,
[2] Linac Coherent Light Source, SLAC National Accelerator Laboratory, Stanford, California 94309, USA,
[3] Department of Applied Physics, Stanford University, Stanford, California 94305, USA

Correspondence: Sina M Koehlenbeck sina.koehlenbeck@stanford.edu Diling Zhu dlzhu@slac.stanford.edu

## Abstract

**The past decades have witnessed the development of new X-ray beam sources with brightness growing at a rate surpassing Moore's law. Current and upcoming diffraction limited and fully coherent X-ray beam sources, including multi-bend achromat based synchrotron sources and high repetition rate X-ray free electron lasers, puts increasingly stringent requirements on stability and accuracy of X-ray optics systems. Parasitic motion errors at sub-micro radian scale in beam transport and beam conditioning optics can lead to significant loss of coherence and brightness delivered from source to experiment. To address this challenge, we incorporated optical metrology based on interferometry and differential wavefront sensing as part of the X-ray optics motion control system. A prototype X-ray optics system was constructed following the optical layout of a tunable X-ray cavity. On-line interferometric metrology enabled dynamical feedback to a motion control system to track and compensate for motion errors. The system achieved sub-microradian scale performance, as multiple optical elements are synchronously and continuously adjusted. This first proof of principle measurement demonstrated both the potential and necessity of incorporating optical metrology as part of the motion control architecture for large scale X-ray optical systems such as monochromators, delay lines, and in particular, X-ray cavity systems to enable the next generation cavity-based X-ray free electron lasers.**

## Introduction

Modern X-ray beam sources and their applications rely heavily on high performance precision motion control mechanisms. State-of-the-art X-ray microscopes now routinely produce tomographs 10 nm spatial resolution in 3D on buried nano structures (1). High resolution X-ray monochromators and analyzers have extended X-ray spectroscopic techniques to unprecedented energy and momentum resolutions (2). Precision metrology and alignment of nano-focusing optics have generated unprecedented intensity levels using X-ray Free Electron Lasers (FELs) and ushered in the era of nonlinear X-ray physics (3) (4). All these developments required precise and stable positioning of optical components with respect to the source, the sample, and other optical components. Our current ability in motion control precision and accuracy in the angular domain for multi-optic systems limits the full realization of the potential brought about by the recent source brightness and coherence improvements (5) (6) (7) (8) (9).

Take the key beam transport and conditioning optics, X-ray monochromators, as an example. X-ray monochromators usually consist of multiple optical elements and require high accuracy coordinated motion to maintain beam position at 10s –100s of meters downstream. Angular deviation from the ideal motion trajectory at a small fraction of the beam divergence scale leads to unacceptable beam motion at the sample position. A key requirement is the stability of the beam during an energy scan, which requires sometimes up to meter-long linear translation of the crystal optics while desiring to limit the angular motion crosstalk to well below 100 nrad scale (10) (11). However, laser interferometry has developed over the years into a reliable tool for precision displacement measurements. The applications range from commercial devices to highly specialized applications in fundamental research such as gravitational wave measurements (12). Laser interferometry can not only serve as an example for the dynamics in a X-ray laser amplifier, but it can also be used as a guidance system for X-ray optics. The key reasons for this are the continuous observation of the phase change of the continuous wave laser source and the resolution accuracy in the picometer and nanoradian range (13) (14).

Intense research and development activities are ramping up around the cavity based FEL concepts (15) (16) (17) and demonstration experiments (18) (19), which promises another 2-3 orders of magnitude improvement in longitudinal coherence and beam brightness over current and planned X-ray FEL sources. Its realization requires stable X-ray cavities to support multi-pass amplification. To realize a continuously-tunable cavity based X-ray laser source, required for nonlinear X-ray spectroscopy, X-ray quantum optics, and quantum metrology (20), cavity design would involve multiple X-ray optical element moving synchronously in coordination while maintaining the overall cavity length and alignment, in order to maintain the lasing interaction between the X-ray pulses and the electron bunches (21).

One of the tunable cavity geometries under consideration is shown in Fig 1, which would be compatible with the planned Linac Coherent Light Source (LCLS)-II HE undulator infrastructure (9). In such tunable cavity, two symmetric groups of optics are laid out at the two opposite ends of the undulators. The downstream optics group introduces a U turn to the X-ray radiation with a relatively small offset, thus limiting the transverse footprint of the optical layout to be compatible with the undulator tunnel layout. The upstream optics group realizes another U turn to send beam back into the undulators. The pairs of Bragg optics can be chosen to fit the desired photon energy with a combined Bragg angle value just below 90 degrees, while the grazing incidence mirror completes the precise 180-degree beam return and introduces the angular, thus photon energy tunability of 10 eV. To operate such a cavity, one would need to perform two basic optical adjustments as indicated in Fig 1. b) and c): cavity length and cavity resonance wavelength adjustment. The motion accuracy requirement is such that during the tuning, the recirculated X-ray pulse, ~50 μm in transverse size, and ~10 μm in length, must maintain overlap with respect to a similarly sized electron bunch traveling along the undulator with a typical length larger than 100 m. This translates to alignment requirement to maintain path length down to ~1 micron level and combined angular alignment down to 100 nrad scale. In this work we present a motion control architecture, implemented on a 3-element X-ray optical configuration that forms half of such tunable X-ray cavity, with the goal of meeting such tight requirement with laser interferometry and active position feedback control.

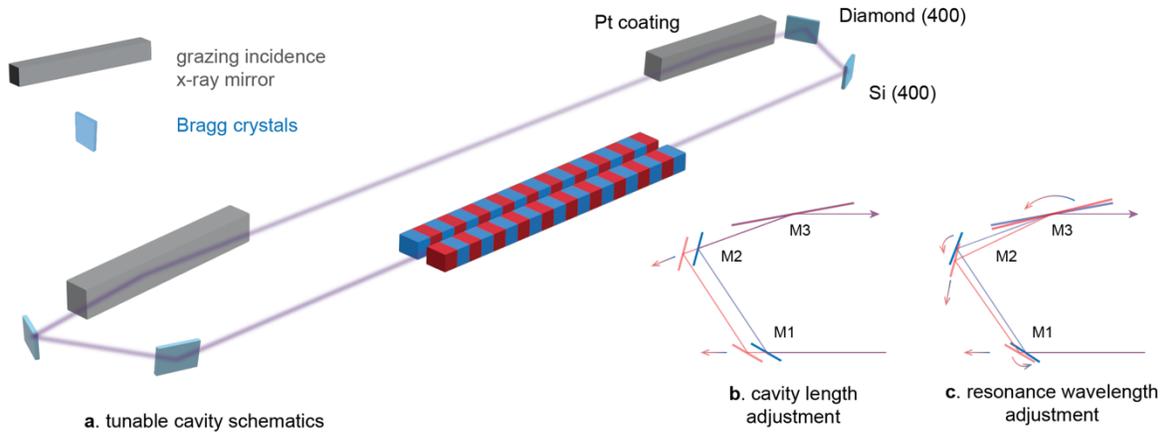

*Figure 1.* ***a***. *Schematics of a 6 optic X-ray cavity optics layout that provides a small range of tunability.* ***b***. *Arrows indicating coordinated motion required for M1 and M2 to extend the cavity length by going from the blue to red beam path, with a fixed input and output beam path.* ***c***. *Arrows indicating coordinate motion required for M1, M2, and M3, to adjust cavity resonance wavelength while maintaining overall cavity length and as well as output beam path.*

Experimental Setup

The 3-element X-ray optics setup was constructed as illustrated in Figure 1 b/c and Fig 2. The optical path is defined by two Bragg crystal reflections and a shallow-angle reflection from a platinum coated X-ray mirror. The Bragg reflections of Si (400) and diamond (400) were chosen such that the cavity resonant wavelength is centered around 8.34 keV, near the Ni K edge. The silicon (400) reflection has a Bragg angle of ~33.2 degrees, and the Diamond (400) reflection ~56.4 degrees. The grazing incidence flat mirror is 300 mm long, with a critical angle of 0.5 degree at the chosen photon energy. At ~0.4 degree incidence angle it makes up the difference to 90 degree total incidence angle, and completes the 180 degree return of the X-ray beam with a horizontal offset of 300 mm. This mimics half of the tunable cavity as shown in Figure 1a. Fig 1b illustrates the simultaneous linear motions required for cavity length adjustment. To mimic photon-energy adjustment, the Bragg condition of the two X-ray crystals must be adjusted simultaneously. The grazing incident mirror will change its angle to maintain the incidence angle sum of exactly 90 degrees. In addition, the exit beam offset, and effective path length shall remain constant via translational motions. In total, three mirrors' rotations plus translations of the two Bragg crystals are needed. This six-axis motion is illustrated in Figure 1c and needs to be executed in coordination.

The test target length scan range is equivalent to a round trip time adjustment of 10 ps. The energy scan covers a range of 10 eV. The motion trajectory for all 3 optics is calculated to maintain the outgoing beam pointing angles. Due to the small relative range, a linear approximation of the calculated motion trajectory is sufficient. For the energy scan, path length is maintained as well. The expected parasitic motion error from the mechanical stages will lead to changes to outgoing beam's position and pointing angle. This is measured using an X-ray profile monitor at 7.2 m distance from the grazing incident X-ray mirror, shown in Figure 2. It images the 'returned' X-ray beam profile for each x-ray pulse at 120 Hz. The X-ray beam is focused on the detector scintillator screen with a compound refractive beryllium lens stack just downstream of the x-ray mirror. This focusing and imaging setup provides an effective angular resolution of 74 nrad/pixel in 7.2 m distance with a pixel size of 0.53 um. The captured single pulse X-ray images are fitted with a 2D Gaussian, with their centers determined down to close to $1/10^{th}$ of a pixel or ~50 nm precision.

Parallel to the X-ray beam, offset in the vertical direction an IR laser beam probes the same mechanical mounts, M1, M2, and M3, as the X-ray beam and is used to measure the integrated angular errors during length and energy adjustment, via laser interferometry. The layout of the measurement interferometer is shown in figure 2a, with its detection highlighted in the blue circle. The laser interferometer measures the variances in length, pitch, and yaw continually. It is routed as close as possible to the X-ray optical element to increase the correlations between the X-ray and laser beam motion. Unintended length changes introduced by the motion trajectories are compensated for by moving M1 and M2 simultaneously, while deviations from the nominal beam pitch and yaw angles are compensated for by tilting M2. A reference interferometer is shown in the yellow circle, which measures length changes that are unrelated to motions of M1, M2 and M3 and are removed by subtraction and feedback control to a fiber stretcher.

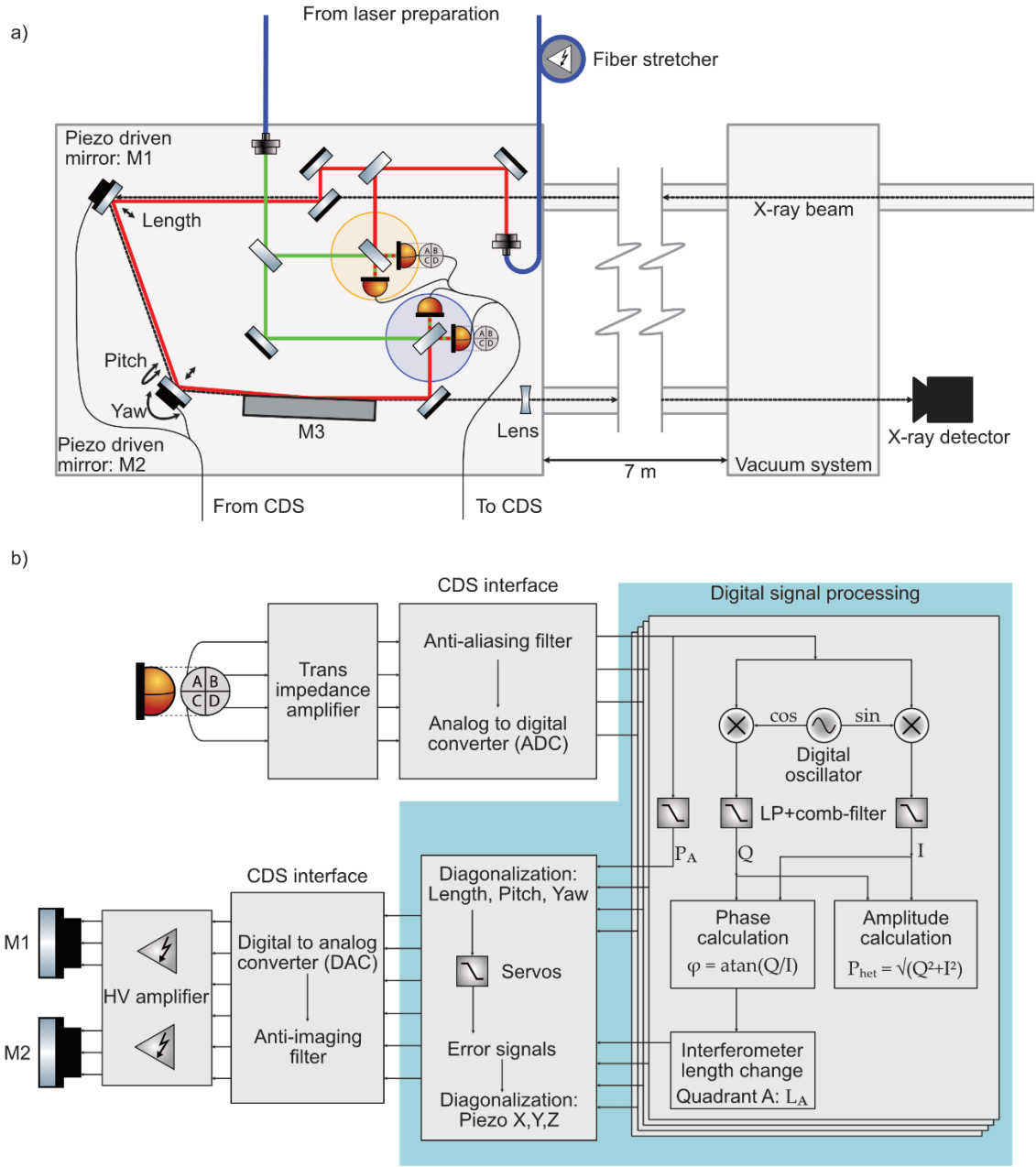

***Figure.2*** *Design of the laser assisted measurement and feedback system **a**. Layout of the X-ray beam path (dashed black line) and the optical guidance system (red and green). The laser light for the interferometric read out is prepared outside of the vacuum chamber and guided via optical fibers into the vacuum system. The two laser beams have a frequency offset of 4096 Hz and hence the interference generates a beat note at this frequency. Two interferometers are formed from them, and their detection is highlighted in circles. The yellow circle shows the detection of a reference interferometer, which is used to suppress common mode phase noise. The actual measurement interferometer is marked with a blue circle. The laser beam, shown in*

*red, samples the X-ray crystal, and mirror assemblies M1, M2 and M3.* **b.** *Schematics of the data acquisition and control flow for the interferometer readout and motion control feedback architecture. All photodiode elements are first converted to voltages with Transimpedance Amplifiers (TIAs) and digitized with an ADC. The signal processing is performed with a LIGO style Control and Data System (CDS). The length changes in the interferometer are measured via phase tracking to a digital oscillator at the beat note frequency and control signals computed and distributed to the piezoelectric mirror mounts.*

The interferometer is read out by heterodyne interferometry. The signal processing chain is sketched in figure 2b. A more detailed description can be found in the methods section and follows the design ideas of the Laser Interferometer Space Antenna (LISA) pathfinder metrology (14) and related experiments (23). The laser interferometer signal is measured with a Si-photodetector and digitized with a Control and Data System (CDS) (24) developed for and by the advanced Laser Interferometer Gravitational-Wave Observatory (aLIGO) (25) collaboration. In the digital domain the interferometer signals are demodulated at the beat frequency and the phase shift over time of this beat note is measured relative to a digital reference sinusoidal signal at the same frequency. The phase shift is calculated to a length shift in the interferometer.

Pitch and yaw are measured with a combined measurement of differential phase shift on a Quadrant Photodiode (QPD) with Differential Wavefront Sensing (DWS) and a simultaneous position measurement on the diode based on the Differential Power Sensing (DPS). The latter one is used to correct the DWS signal from the residual introduced through the radius of curvature of the wavefront (26). The corresponding equations can be found in the methods section. The interferometer read-out jitter has been measured to be 14 nm in length, 72 nrad in pitch and 21 nrad in yaw for the static case.

The information from the interferometer is processed in real time with the CDS, which computes control signals with unity gain frequencies of $10^{th}$ of Hertz. The error signals are converted to analog voltages, amplified, used to drive the two 3-axis Piezo mounts. The actuation onto the compound mirror assemblies has been diagonalized to the degrees of freedom, length, pitch, and yaw with a second interferometer in place of the X-ray beam. The control system was originally designed to control 3 degrees of freedom but was extended to 5 degrees of freedom to achieve a further reduction in movement by including translational degrees of freedom.

## Results

The setup was installed, and the measurements performed at the X-ray Pump Probe instrument at the Linac Coherence Light Source (27). The performance of the system is evaluated by direct measurement of the transverse x-ray beam position at the profile monitor. With the control feedback loop off, and all motion axis static, a shot-to-shot variance of the X-ray beam position indicated angular RMS pointing jitter of 0.459(3) µrad in pitch and 0.169(1) µrad in yaw, deduced based on the 7.2m baseline distance of the x-ray focusing lens. This assumes that the position jitters primarily are contributed from pointing errors. This sets the noise floor for the x-ray measurement.

Beam pointing errors measured during two types of delay scans and one energy scan are shown in Figure 3 for both pitch (a) and yaw (b). The red data points represent the beam angle deviations from the nominal set points during the delay and energy scan without any interferometer feedback. The same holds true for the blue data points, but this time the feedback control system is actively correcting for the deviations from the set-point. The black line in all graphs of figure 3 is the moving median of 250 measurement points. The spread around the median is the shot-to-shot jitter of the X-ray beam source, and from up-stream optical elements. From the data in figure 3 two important quantities can be analyzed, the shot-to-shot variance and the peak-to-peak variation of the moving median. The first one is a diagnostic, to whether angular vibrations from the stages during the coordinated movement have been imprinted onto the X-ray beam or if the feedback system introduces errors. The latter represents the absolute deviation from the nominal position and a measure for misalignment.

To analyze the shot-to-shot jitter the moving median has been removed from the data points and the results for the data sets during active feedback are shown in figure 4b-d. Figure 4a shows the shot-to-shot jitter of the X-ray beam source in the static case. All results are also listed for comparison in table 1 for pitch and table 2 for yaw. For the short delay and energy scan the jitter is equivalent to the static case and the shot-to-shot jitter during the open loop scan. However, the long delay scan shows a splitting distribution, best understood from the histogram in figure 4b. This is a systematic artifact introduced by spurious interferometers due to changes in absolute length of the interferometer (28). It could be minimized by an alignment on the diode to balance the effect equally on all four quadrants, however during this experimental time, the interferometer accumulated an alignment offset and the error became apparent. Remote balancing is an option to reduce the error in future iterations of the experiment and can be removed on the length measurement with balanced detection (29).

The feedback control system has in all three scans reduced the peak-to-peak variance of the moving median. The alignment error (peak-to-peak) during the long delay scan has been reduced from 4.09 µrad to 0.62 µrad in pitch and from 10.1 µrad to 1.13 µrad in yaw. During the closed loop energy scan the error was 1.23 µrad and 1.76 µrad for pitch and yaw respectively. The short delay scan shows exceptional stability, which is equivalent to the static case in short term jitter. This is a result of a slightly different controls diagram implementation. Instead of correcting the DWS signal with the DPS signal to compensate for pitch and yaw, the DPS signal itself was used in a feedback loop to control M1 pitch and yaw. This resulted in a stable output beam in tilt and displacement with a peak-to-peak median of 0.234 µrad in pitch and 0.15 µrad in yaw. However, the feedback to M1 and M2 was not diagonalized. This leads to a Bragg condition misalignment between M1 and M2 beyond the 2.5 ps range. The result, however, demonstrated the potential that with feedback activated in 5 degrees of freedom, the output beam can be stabilized down to the scale of 100 nrad.

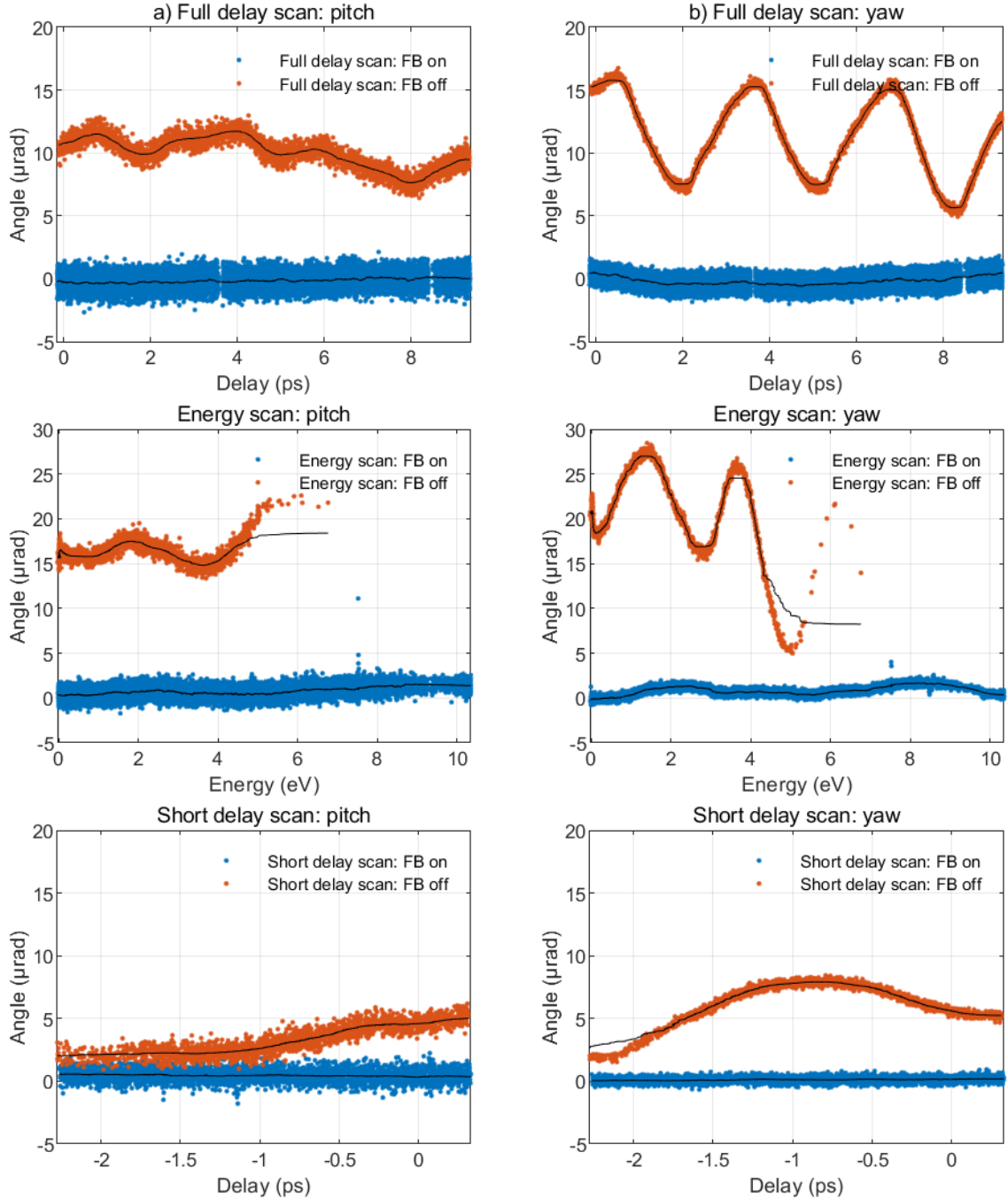

*Figure 3. Measured center of the X-ray beam on the detector in units of equivalent beam tilt. The red data points were taken during delay and energy scans without feedback control form the interferometer and blue show the equivalent scans with the feedback enabled. Panel a. shows pitch (vertical tilt), and b. yaw (horizontal tilt) during a short and long delay scan and an energy scan of the halve cavity experiment. The black lines are the moving median average of 250 data points. A reduction in peak-to-peak motion once the feedback loops are active can be seen.*

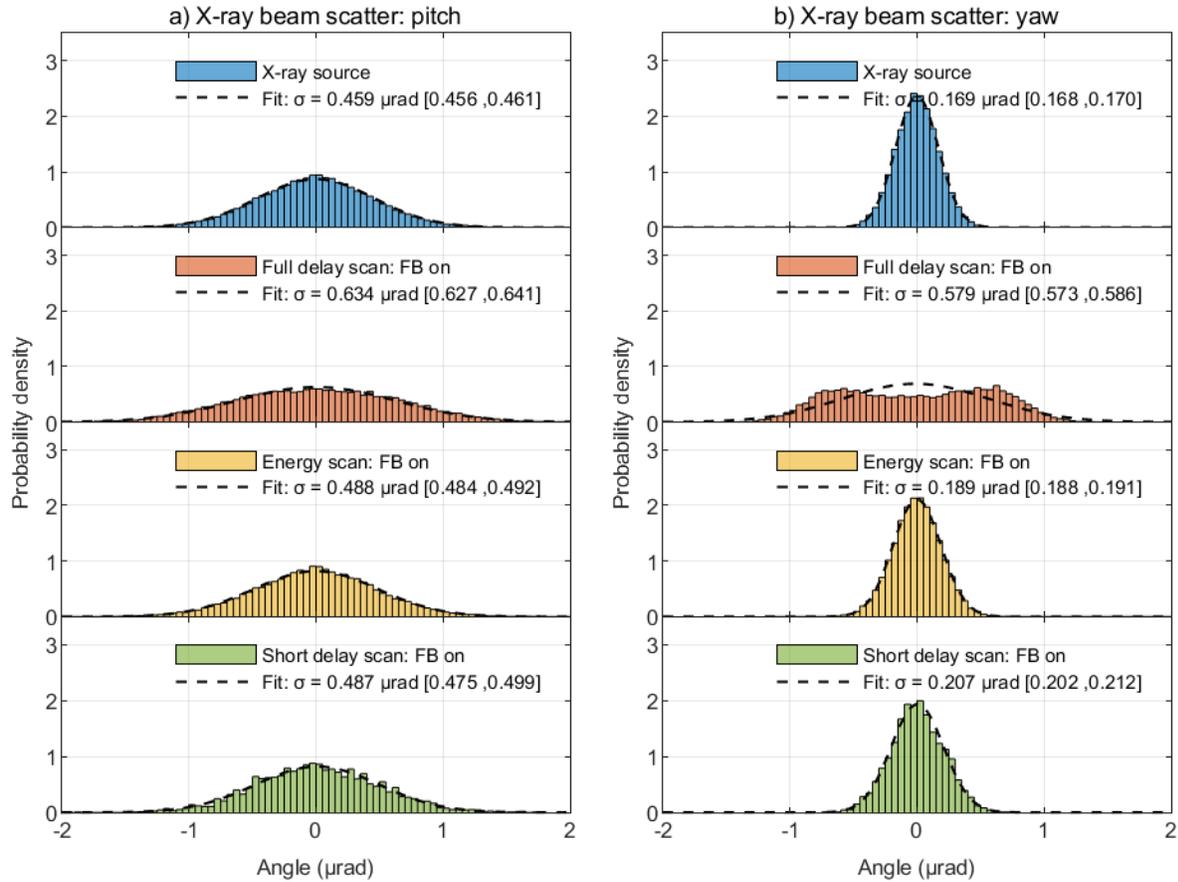

*Figure 4: Analysis of the shot-to-shot variance of the X-ray beam source tilt in the static case (blue) and during the delay scan (red), the energy scan (yellow) and a short delay scan with feedback in 5 degrees of freedom (green). The moving median shown in Figure 3 was subtracted from the data so that the remaining variance corresponds to the shot-to-shot beam jitter. Panel a. shows the beam pitch and panel b. shows the yaw of the beam. Comparing the delay and energy scans with the reference data in blue, the feedback control introduces no jitter, except in the case of the full delay scan. A splitting of the data points can be seen most clearly in the yaw and results from the systematic error caused by spurious interference.*

*Table 1) Measured errors during the energy and delay scans, as well as the static reference for the beam pitch of the x-ray beam.*

| a) Pitch error | Scatter: sigma [95% confidence interval] (µrad) | | RMS (µrad) | | Moving median: peak-peak (µrad) | |
|---|---|---|---|---|---|---|
| Feedback | Off | On | Off | On | Off | On |
| Static reference | 0.459(3) | - | 0.535 | - | 0.668 | - |
| Short delay scan | 0.446(14) | 0.487(12) | 1.88 | 0.657 | 2.81 | 0.234 |
| Full delay scan | 0.464(10) | 0.634(7) | 1.27 | 0.664 | 4.09 | 0.62 |
| Energy scan | 0.509(14) | 0.488(4) | 1.6 | 0.861 | 3.36 | 1.23 |

*Table 2) Measured errors during the energy and delay scans, as well as the static reference for the beam yaw of the x-ray beam.*

| b) Yaw error | Scatter: sigma [95% confidence interval] (µrad) | | RMS (µrad) | | Moving median: peak-peak (µrad) | |
|---|---|---|---|---|---|---|
| Feedback | Off | On | Off | On | Off | On |
| Static reference | 0.169(1) | - | 0.331 | - | 0.162 | - |
| Short delay scan | 0.248(8) | 0.207(5) | 1.93 | 0.238 | 4.93 | 0.15 |
| Full delay scan | 0.236(6) | 0.579(7) | 4.98 | 0.669 | 10.1 | 1.13 |
| Energy scan | 0.393(13) | 0.189(2) | 4.23 | 0.877 | 17.5 | 1.76 |

## Discussion

The demonstrated performance of X-ray beam path stabilization as shown in the short-range delay scan at < 0.25 µrad peak-to-peak error, satisfies what is required to establish temporal (longitudinal) overlap, after transverse overlap is established. In such an alignment process, the transverse alignment would be first established by recirculating a single X-ray pulse and making sure the X-ray enters and exits the undulator following the same beam path as the electron beam (30). This will be then followed by turning on the high repetition rate electron bunch train and performing cavity length adjustment from a starting length defined by metrology with an error of $10^{-6}$, or e.g. ±1 ps assuming 1 microsecond bunch spacing at 1 MHz repetition rate from the accelerator. It is also encouraging to see that with relatively basic stepper motor driven motion stages, continuous cavity length tuning can be executed without introducing excessive pointing jitter beyond the natural pointing jitter of the SASE FEL itself.

The controls loop encountered challenges in operating with the same 5-axis feedback loop for the longer-range delay scan and the energy scan. A cross-coupling on feedback between M1 and M2 angle led to diverging angular corrections, such that the two Bragg crystals became misaligned from each other. The control system was originally designed to control 3 degrees of freedom and it was implemented for the longer-range scans. It addressed pointing errors but showed residue in beam displacement errors. The output pointing error of 1-2 µrad peak-to-peak is already close to what would be required for realizing the high-gain cavity system, as the photon-to-electron interaction in the undulator leads to a strong optical and gain guiding effects and thus reduces the tolerable mirror error to 0.5 µrad RMS (15). We note that the performance of the interferometric measurement has been compromised during a particular part of the experiment, but this can be resolved for the length measurement with balanced detection and with alignment to the QPD for the DWS measurement. There is a clear path to extend the performance as demonstrated in the short delay scan at <0.25 µrad level to larger range and more complex adjustments ranges by design adjustment of the controls loop. The low gain X-ray FEL oscillator requirement is within reach with the presented methodology.

We envision the measurement system presented here as an effective motion error tracking and compensation mechanism for co-located optical components such as in x-ray monochromators, delay lines, and mirror systems. For realizing a large-scale cavity as described in Fig 1 a), the motion guiding system can be expanded to include two 'local' tracking systems to secure motion accuracy of the components within individual optical bench on each end. Another single beam 'distance + angle' interferometer can be used to connect the two optical benches separated by large distances to make the two benches virtually 'monolithic and stable'. Such an architecture utilizing relatively mature optical interferometer methodology can be essential in realizing future cavity-based x-ray free electron lasers driven by high repetition rate accelerators.

## Materials and methods

The following section describes the details of experimental set up, data acquisition system, and data processing protocol.

### Experiment

The experiment was performed at the X-ray Pump Probe (XPP) (27) instrument at LCLS. The photon energy of the FEL was calibrated near the absorption edge of Ni K shell and set to 8.34 keV, with a bandwidth of ~ 30eV full width. A float zone Si(400) crystal and a synthetic single crystal diamond (400) crystal were chosen as the Bragg optics. The grazing incidence mirror has a length of 300mm and was coated with Pt. The cavity was initially aligned with monochromatic beam at 8.350 keV for calibration (11). At this photon energy the Bragg angle of the Si crystal M1 is 33.14 deg; the diamond crystal M2 has a Bragg angle of 56.36 deg. The starting angle of the grazing incidence mirror M3 is at 0.5 deg. From this starting point the energy scans were executed in the negative direction from 8.35keV to 8.34keV.

The Bragg crystals are mounted onto 3-axis piezo mirror mounts (NewportPSM2SG-D) that hosts both the x-ray crystal and the mirror for the interferometer IR beam. The piezo motors are mounted on stepper motor driven goniometer stage stack with larger range rotational motions to steer the beam reflection in both horizontal (yaw, $\theta_{1,2}$) and vertical (pitch, $\chi_{1,2}$). In addition M1 can translate in in z direction ($z_1$) along the incoming beam direction. M2 can translate in both the beam direction ($z_2$) and the transverse horizontal direction ($x_2$). The grazing incident mirror M3 has stepper motor driven motion to adjust the incidence angel ($\theta_3$), and translate in transverse horizontal direction ($x_3$). Pre-programmed motion trajectory executing an Energy and Delay scan have the following motion vectors:

|  | $\Delta z_1$ (mm) | $\Delta z_2$ (mm) | $\Delta x_2$ (mm) | $\Delta \theta_1$ (deg) | $\Delta \theta_2$ (deg) | $\Delta \theta_3$ (deg) | $\Delta E$ (eV) | $\Delta t$ (ps) |
|---|---|---|---|---|---|---|---|---|
| **Energy scan** | -1.451 | -1.286 | 2.946 | 0.0906 | 0.3906 | 0.3 | 21.42 | 0 |
| **Delay scan** | 2 | 1.992 | -0.0174 | 0 | 0 | 0 | 0 | 13.34 |

The laser interferometer is routed parallel but above the X-ray beam, at an elevation 25mm higher. It's in plane angle closely follows that of the X-ray except a higher grazing angle was used on the grazing incidence x-ray mirror M3 by 0.5 to reduce clipping by the mirror edges. The optical mirrors and x-ray crystals on M1/M2 motion stack were rigidly connected mechanically to minimize angular relative drift. The layout of the laser interferometer is shown in figure 3a. It is in a heterodyne Mach-Zehnder configuration operating at wavelength of $\lambda = 1064$ nm. The optical beams are in-coupled into the vacuum vessel via polarization preserving optical fibers and couplers. For the heterodyne interferometry the laser light from one fiber has a frequency offset of 4096 Hz as compared to the light from the other fiber. When the two beams are merged and interfered with one another, a beat note at the heterodyne frequency is observed. A single element diode is used to measure the length change $\Delta l$ of the interferometer from the phase change $\Delta \varphi$ of the interferometric beat note as

$$\Delta l = \frac{\lambda}{2\pi} \cdot \Delta \varphi.$$

The data processing in the CDS is illustrated in figure 2b and based on the digital two quadrature demodulation described in ref. (31).

The beam delivery via two optical fibers can introduce differential phase noise into the measurement system, which was tracked and compensated. In figure 2a the optical layout of the laser measurement system is shown. The yellow and blue circle highlight the detection of the two interferometers, in yellow the reference interferometer and in blue the measurement interferometer. The reference interferometer is used to measure phase noise of the laser preparation and delivery system. This measurement is subtracted from the measurement of the blue interferometer to remove common mode phase noise. In addition, a fiber stretcher is used to cancel the differential pathlength noise in a feedback control loop.

The beam pointing deviation is measured via DWS and DPS with a QPD. DPS is the normalized position measurement on a segmented photodiode and measured as

$$\mathrm{DPS}_{x,y} = \frac{P_{\mathrm{right,top}} - P_{\mathrm{left,bottom}}}{\sum P}.$$

Calibration to position in $x$ and $y$ on the photodiode is related to the beam size $\omega_m$ of the measurement beam on the diode as well as the power $P_{m,r}$ of the measurement and reference beam respectively. Following (32) the position on the diode is measured to be

$$x = \sqrt{\frac{\pi}{8}} \cdot \omega_m \cdot \frac{P_m + P_r}{P_m} \cdot \mathrm{DPS}_x,$$

$$y = \sqrt{\frac{\pi}{8}} \cdot \omega_m \cdot \frac{P_m + P_r}{P_m} \cdot \mathrm{DPS}_y.$$

The calibration factor is

$$c_{\text{DPS}} = \sqrt{\frac{\pi}{8}} \cdot \omega_m \cdot \frac{P_m + P_r}{P_m}.$$

The DWS signal is measured in relation to the interferometric phase measurement $\varphi$ as

$$\text{DWS}_{\varphi,\text{yaw,pitch}} = {}^1\!/_2\, \varphi_{\text{right,top}} - {}^1\!/_2\, \varphi_{\text{left,bottom}}.$$

Following (26), the tilt $\alpha$ of the wavefront is related to the DWS measurement in yaw as

$$\text{DWS}_{\varphi,\text{yaw}}(\alpha, x) = \sqrt{\frac{2}{\pi}} \cdot k \cdot \omega_{\text{eff}} \cdot \left(\alpha - \frac{x}{R_m}\right) \cdot F(\sigma) + O(\alpha^2, x^2),$$

With the wavenumber $k$, the effective beam radius $\omega_{\text{eff}}$, the beam tilt against the reference beam $\alpha$, the horizontal displacement on the QPD $x$, the radius of curvature of the wavefront of the measurement beam $R_m$, and

$$F(\sigma) = \frac{1}{\sqrt{2}} \sqrt{\frac{1 + \sqrt{1 + \sigma^2}}{1 + \sigma^2}}, \sigma = \frac{k\omega_{\text{eff}}^2}{4R_{\text{rel}}}, \frac{1}{R_{\text{rel}}} = \frac{1}{R_r} - \frac{1}{R_m}, \frac{1}{\omega_{\text{eff}}^2} = \frac{1}{\omega_r^2} - \frac{1}{\omega_m^2}.$$

$R_{r,m}$ and $\omega_{r,m}$ are the radius of curvature and beam radius at the QPD of the reference and measurement beam respectively. The equivalent equation for pitch and vertical displacement holds true. The DWS signal can be corrected with the DPS signal to turn it into a pure angular measurement. We calibrated the DPS correction signal by displacing the laser beam on the diode by $x$ in the horizontal direction, without introducing a tilt $\alpha$, which results in.

$$\text{DWS}_{\varphi,\text{yaw}}(x) = -\sqrt{\frac{2}{\pi}} \cdot k \cdot \omega_{\text{eff}} \cdot \frac{x}{R_m} \cdot F(\sigma).$$

This is subtracted and the remaining DWS signal only depends on the angle $\alpha$

$$\text{DWS}_{\varphi,\text{yaw}}(\alpha) = \sqrt{\frac{2}{\pi}} \cdot k \cdot \omega_{\text{eff}} \cdot \alpha \cdot F(\sigma).$$

The angle is measured as

$$\alpha = \sqrt{\frac{\pi}{2}} \cdot \frac{1}{k \cdot \omega_{\text{eff}} \cdot F(\sigma)} \cdot \text{DWS}_{\varphi,\text{yaw}},$$

In our signal processing chain, the phase is converted to an equivalent length change before the DWS signal is calculated, hence $k \cdot \text{DWS}_{l,\text{yaw,pitch}} = \text{DWS}_{\varphi,\text{yaw,pitch}}$. The angle measurement in our signal processing chain is calculated as

$$\alpha = \sqrt{\frac{\pi}{2}} \cdot \frac{1}{\omega_{\text{eff}} \cdot F(\sigma)} \cdot \text{DWS}_{l,\text{yaw}},$$

$$\beta = \sqrt{\frac{\pi}{2}} \cdot \frac{1}{\omega_{\text{eff}} \cdot F(\sigma)} \cdot \text{DWS}_{l,\text{pitch}}.$$

The calibration factor is

$$c_{\text{DWS},l} = \sqrt{\frac{\pi}{2}} \cdot \frac{1}{\omega_{\text{eff}} \cdot F(\sigma)}.$$

All calibration factors have been determined experimentally by rotation and translation of the motion stages.

The beam tilts $\Delta\alpha$ and $\Delta\beta$ and length change $\Delta l$ of the interferometer are related to the coordinate system of the half-cavity experiment, with the mirror tilts M2 yaw ($\Delta\theta$) and pitch($\Delta\chi$) and the cavity length $\Delta z$ as follows:

$$\Delta z = \tfrac{1}{2} \cdot \Delta l$$

$$\Delta\theta = \tfrac{1}{2} \cdot \Delta\alpha$$

$$\Delta\chi = \tfrac{1}{2} \cdot \Delta\beta$$

The data is presented in the coordinate of beam pitch ($\Delta\beta$) and yaw ($\Delta\alpha$) projected to mirror M2 and interferometer length change ($\Delta l$). The X-ray diagnostics does not allow us to measure the length, only the beam tilts.

### Data acquisition and processing

The data processing is divided into two independent data acquisition and control systems.

The LCLS data acquisition (DAQ) system records the data stream related to the X-ray diagnostics, the accelerator condition, as well as the beamline status. It is event-driven, synchronized to the generation of X-ray pulses on a 120 Hz clock. The images from the X-ray camera are captured at this rate. Post processing reduces the data to smaller packets containing fitted beam x-centroid and y-centroid, in synchronized array format with other beamline data such as pulse intensities, motor positions, accelerator parameters, etc. The synchronous motion control was executed using built in functions of the Aerotech Ensemble multi-axis motion controller. An analog signal linear to position within the planned motion trajectory was generated and sent from the controller to the analogue input of the LCLS DAQ. This synchronized ADC value can then be used to calculate the equivalent energy change and time delay of a cavity, for each recorded x-ray beam image.

Independent of the LCLS DAQ, a second data acquisition stream runs in parallel via the LIGO CDS (24) and is responsible for the interferometer data processing and the real-time control loops. The sampling rate is 65536 Hz, but the data is filtered and stored as proprietary frame files at a rate of 2048 Hz. This data acquisition is not synchronized to the LCLS data acquisition system, but to the Network Timing Protocol (NTP) and therefore to GPS time. Software tools allow quick and easy access to stored and live data. We have exported selected frame data to mat files.

Data processing

The X-ray and interferometer data were then analyzed in MATLAB. As part of the X-ray data set, the intensity on the detector was recorded and a threshold was used for the X-ray data to exclude weak X-ray images without sufficient signal-to-noise ratio for an accurate centroid position. The centroid data is converted to an equivalent tilt angle of the beam. To measure shot-to-shot variance over the course of the experiment, a moving median of 250 data points was calculated and subtracted from the raw data. Large-scale movements are suppressed and the short-term jitter of the X-ray beam on the detector is shown in the histograms in Figure 4, normalized to the probability density function (pdf). A fit to the normal distribution gives the standard deviation and is shown in Table 1. Data points had to be excluded for the shot-to-shot variance calculation of the open-loop energy scan. The outliers result from the deviation of the moving median from the data points. To calculate a histogram of the shot-to-shot jitter, 250 data points are required. The speed of the scan resulted in relatively few data points per energy interval and in addition the Bragg reflection condition was lost, and the number of data points further reduced. This would artificially increase the shot-to-shot jitter and therefore only data from the scan that was not too far from the median was used.

## Acknowledgements

The authors would like to thank Edgard Bonilla, Johannes Lehmann and Gerhard Heinzel for their views and discussions. The authors acknowledge support from the Laboratory Directed Research and Development program at the SLAC National Accelerator Laboratory and the National Science Foundation under Grant No. NSF-2011786. Use of the Linac Coherent Light Source (LCLS), SLAC National Accelerator Laboratory, is supported by the U.S. Department of Energy, Office of Science, Office of Basic Energy Sciences under Contract No. DE-AC02-76SF00515.

17. *Active Q-Switched X-Ray Regenerative Amplifier Free-Electron Laser.* **Tang, Jingyi, et al.** s.l. : American Physical Society (APS), July 2023, Physical Review Letters, Vol. 131, p. 055001. ISSN: 1079-7114.

18. *Low-loss stable storage of 1.2 Å X-ray pulses in a 14 m Bragg cavity.* **Margraf, Rachel, et al.** s.l. : Springer Science and Business Media LLC, August 2023, Nature Photonics, Vol. 17, pp. 878–882. ISSN: 1749-4893.

19. *Cavity based x-ray free electron laser demonstrator at the European X-ray Free Electron Laser facility.* **Rauer, Patrick, et al.** s.l. : American Physical Society (APS), February 2023, Physical Review Accelerators and Beams, Vol. 26, p. 020701. ISSN: 2469-9888.

20. **Adams, Bernhard, et al.** Scientific Opportunities with an X-ray Free-Electron Laser Oscillator. *Scientific Opportunities with an X-ray Free-Electron Laser Oscillator.* s.l. : arXiv, 2019.

21. *Tunable optical cavity for an x-ray free-electron-laser oscillator.* **Kim, Kwang-Je and Shvyd'ko, Yuri V.** s.l. : American Physical Society (APS), March 2009, Physical Review Special Topics - Accelerators and Beams, Vol. 12, p. 030703. ISSN: 1098-4402.

22. *Cavity-Based Free-Electron Laser Research and Development: A Joint Argonne National Laboratory and SLAC National Laboratory Collaboration.* **Marcus, Gabriel, et al.** s.l. : JACoW Publishing, Geneva, Switzerland, 2019, Proceedings of the 39th Free Electron Laser Conference, Vol. FEL2019, p. Germany.

23. *A study on motion reduction for suspended platforms used in gravitational wave detectors.* **Koehlenbeck, Sina M., et al.** s.l. : Springer Science and Business Media LLC, February 2023, Scientific Reports, Vol. 13. ISSN: 2045-2322.

24. *advligorts: The Advanced LIGO real-time digital control and data acquisition system.* **Bork, Rolf, et al.** s.l. : Elsevier BV, January 2021, SoftwareX, Vol. 13, p. 100619. ISSN: 2352-7110.

25. *Advanced LIGO.* **Aasi, J., et al.** s.l. : IOP Publishing, March 2015, Classical and Quantum Gravity, Vol. 32, p. 074001.

26. *Measurement of the absolute wavefront curvature radius in a heterodyne interferometer.* **Hechenblaikner, Gerald.** s.l. : The Optical Society, August 2010, Journal of the Optical Society of America A, Vol. 27, p. 2078.

27. *The X-ray Pump–Probe instrument at the Linac Coherent Light Source.* **Chollet, Matthieu, et al.** s.l. : International Union of Crystallography (IUCr), April 2015, Journal of Synchrotron Radiation, Vol. 22, pp. 503–507. ISSN: 1600-5775.

28. *Investigation and compensation of the nonlinearity of heterodyne interferometers.* **Hou, Wenmei and Wilkening, Günter.** s.l. : Elsevier BV, April 1992, Precision Engineering, Vol. 14, pp. 91–98. ISSN: 0141-6359.

29. *Relative-Intensity-Noise Coupling in Heterodyne Interferometers.* **Wissel, Lennart, et al.** s.l. : American Physical Society (APS), February 2022, Physical Review Applied, Vol. 17.
17